\documentclass[pre,twocolumn,aps,eqsecnum]{revtex4}
\usepackage{epsfig}

\begin{document}

\title{Thermal Effects in Dislocation Theory}

\author{J.S. Langer}
\affiliation{Department of Physics, University of California, Santa Barbara, CA  93106-9530}

\date{\today}

\begin{abstract}
The mechanical behaviors of polycrystalline solids are determined by the interplay between phenomena governed by two different thermodynamic temperatures: the configurational effective temperature that controls the density of dislocations, and the ordinary kinetic-vibrational temperature that controls activated depinning mechanisms and thus deformation rates.  This paper contains a review of the effective-temperature theory and its relation to conventional dislocation theories.  It includes a simple illustration of how these two thermal effects can combine to produce a predictive theory of spatial heterogeneities such as shear-banding instabilities.  Its main message is a plea that conventional dislocation theories be reformulated in a thermodynamically consistent way so that the vast array of observed behaviors can be understood systematically.  

\end{abstract}

\maketitle

\section{Introduction}
\label{Intro}
The mechanical properties of solids -- their deformability and modes of failure -- are surely among the central issues in materials research.  After more than half a century of work in this field, however, materials scientists still have no first-principles, predictive theory of phenomena as basic as dislocation-induced strain hardening or fracture toughness. My purpose here is to present a critical analysis of this situation and to argue that recent developments provide an opportunity for significant progress. 

Dislocation-mediated plastic deformations are enormously complex phenomena when observed in microscopic detail. The dislocations are line defects of various kinds that are driven to move through polycrystalline solids in complicated ways as they produce irreversible structural changes.  Their motions are impeded by crystalline defects, grain boundaries and, especially, by other dislocations.  In two recent papers \cite{LBL10-10,JSL-15}  (LBL10 and L15), my colleagues and I have shown that we can solve several of the most important outstanding dislocation problems without making direct reference to most of these complications.  In fact, however, these complications are underlying ingredients of our calculations; but, by starting with basic ideas such as scaling and dimensional analysis, and using the principles of statistical thermodynamics, we have had no need to model those mechanisms in detail in order to understand many of the main phenomena.  

My topic in this paper is thermal effects.  I start by reviewing the elementary but currently unconventional idea that, when driven out of mechanical equilibrium, the configurational degrees of freedom of a solid generally acquire an effective temperature that differs from the ambient temperature.  Just understanding the meaning of this effective temperature is enough to tell us that correct theories of solid deformation must differ from those that continue to be taken seriously in this field. 

The effective temperature provides basic information about the density of structural flow defects, including both dislocations in crystals and shear-transformation zones (STZ's) in amorphous materials. \cite{FL-11} It is relevant to situations in which the solid is deformed by external forces so that energy and entropy flow through it, and so that the population of flow defects falls out of equilibrium with the thermal background.  Ordinary thermal fluctuations are too weak by many orders of magnitude to create or annihilate dislocations, or even STZ's.  Nevertheless, they sensitively determine the rates at which dislocations are depinned from each other or from structural defects, and thus they are especially effective in controlling rates of deformation. 

In what follows, I explore ways in which these two different kinds of thermodynamic effects combine to determine the nonequilibrium dynamics of solids. I focus on the interplay between dislocation induced hardening and thermally induced softening that produces shear-banding instabilities and the like.  In this process, I challenge some long-standing assumptions in this field. 

\section{The Effective Temperature}
\label{Teff}

So far as I can see, there is nothing hypothetical or controversial about the effective temperature, especially not for a deforming solid with dislocations.  To an extremely good approximation, the atomic configurational degrees of freedom of a polycrystalline solid are decoupled from the kinetic-vibrational degrees of freedom.  These two subsystems do exchange energy with each other when groups of atoms undergo irreversible configurational rearrangements.  But, unless the solid is near its melting point, those rearrangements are extremely infrequent on microscopic time scales.  Thus, it is essential to begin any theoretical description of such a system by focusing on the configurational part.  (See \cite{BL-I-II-III-09} for a more systematic discussion than is presented here.)    

For the moment, consider only steady-state, dislocation-mediated shear flow. Also assume that, on average, we need to consider only edge dislocations moving on simple glide planes.   Denote the energy of the configurational subsystem by $U_C(S_C,\rho)$.  Here, $\rho$ is the areal density of dislocations or, alternatively, the total length of dislocation lines per unit volume; and $S_C(U_C,\rho)$ is the  entropy computed by counting the number of atomic configurations, including the number of arrangements of dislocations, at fixed values of $U_C$ and $\rho$.  

The dislocations are driven by external forces to undergo chaotic motions; thus they explore  statistically significant parts of their configuration spaces.  According to Gibbs, this configurational subsystem must be maximizing its entropy $S_C$; that is, it must be at or near its state of maximum probability.  It is doing this at a value of the energy $U_C$ that is determined by the balance between the input power and the rate at which energy is dissipated to the kinetic-vibrational subsystem, which serves here as the thermal reservoir.  The method of Lagrange multipliers tells us to find this most probable state by maximizing the function $S_C - X U_C$, and then finding the value of the multiplier $X$ for which $U_C$ has the desired value.  Define $X$ to be proportional to the inverse of the effective temperature $T_{e\!f\!f}$, i.e. $1/X = k_B T_{e\!f\!f}\equiv\chi$.   Thus, the system finds a minimum of the free energy 
\begin{equation}
\label{Fdef}
F_C = U_C - \chi\,S_C, 
\end{equation}
and
\begin{equation} 
\label{chiUS}
\chi = {\partial U_C\over\partial S_C} .   
\end{equation}

We already can draw some interesting conclusions.  First, note that minimizing $F_C$ in Eqs.~(\ref{Fdef}) and (\ref{chiUS}) determines the steady-state dislocation density $\rho_{ss}$ as a function of the steady-state effective temperature $\chi_{ss}$.  For example, in the simplest approximation, the $\rho$ dependence of $U_C/V$ has the form $e_D\, \rho$, where $V$ is the volume and $e_D$ is a characteristic energy of a dislocation.  Similarly, the $\rho$ dependence of $S_C/V$ has the form $-\,\rho\,\ln\rho + \rho$.  Thus, minimizing $F_C$ produces the usual Boltzmann formula, $\rho_{ss} \propto \exp\,(-\,e_D/\chi_{ss})$.  An appreciable density of dislocations requires a value of $\chi_{ss}$ that is comparable to $e_D$, which is enormously larger than the ambient thermal energy $k_B\,T$.  Moreover, this expression for $\rho_{ss}$ is an increasing function of $\chi_{ss}$.  At larger densities, the dislocations become increasingly entangled with each other, and the system becomes harder in the sense that the stress required for deformation becomes larger. Thus, this rudimentary statistical argument immediately predicts strain hardening.  

Next, note that $\chi_{ss}$ is a measure of the configurational disorder in the material, in direct analogy to the way in which the ordinary temperature determines the strength of energy and density fluctuations.  As such, $\chi_{ss}$ must be a function primarily of the plastic shear rate $\dot\epsilon^{pl}$, which we can think of as the rate at which the system is being ``stirred,'' i.e. the rate at which the atoms are being caused to undergo rearrangements.  If this shear rate is so slow that the system relaxes between rearrangement events, then the steady state of disorder is determined only by the number of atomic rearrangements and not by the rate at which they occur.  That is, $\chi_{ss}$ must be some nonzero constant below a characteristic shear rate whose value is determined by atomic time scales.  It follows that $\rho_{ss}$ is also a constant and that the steady-state stress increases only very slowly over a wide range of low to moderate shear rates.  This is what is observed experimentally. 

This situation changes when the system is being driven fast enough that it does not have time to relax between rearrangement events.  Under those conditions,  $\chi_{ss}$, $\rho_{ss}$, and the driving stress all increase with increasing shear rate.  In LBL10, we showed that a simple, quantitative version of these arguments, using just a few physically motivated parameters, produces an accurate fit to curves of stress versus strain rate for Cu at temperatures $100\,K$, $300\,K$, and $1173\,K$, for shear rates starting at $10^{-3}\,s^{-1}$ and going all the way up through the strong-shock regime to $10^{12}\,s^{-1}$.  The theory that produces these results is based directly on the effective-temperature analysis outlined above.  It adds only the thermally activated depinning mechanism to be described in the next section of this paper, and essentially nothing else.

In view of these arguments and of other calculations reported in LBL10 and L15, it seems to me that attempts to develop a theory of dislocation-mediated plasticity without using the effective temperature are as futile as trying to describe the behavior of simple gases without using the ordinary temperature.  Yet this is exactly what has been done by mainstream dislocation theorists ever since the ground-breaking work of Taylor and Orowan in the 1930's. For example, see recent reviews by Armstrong et. al. \cite{ARMSTRONGetal-09} and Gray \cite{GRAY-12}. The lack of a thermodynamic foundation for their analyses has required these theorists to make a wide range of phenomenological assumptions.

For example, it is generally asserted in the dislocation-theory literature that flow stresses at high strain rates are largely due to viscous forces impeding the motions of free dislocations.  So far as I know, however, there never has been a careful test of this assertion, nor any direct experimental observation to support it.  In LBL10, we argued that viscous forces could be important only in the opposite limit of small strain rates and small dislocation densities, where the time taken for a freely moving dislocation to traverse the distance between pinning sites might be greater than the pinning time.  At higher strain rates, we showed that the effective-temperature analysis explains the observed behaviors naturally and simply. 

Another example that I have found especially interesting is the rate-hardening anomaly in Cu described by Follansbee and Kocks in 1988. \cite{FOLLANSBEE-KOCKS-88} Here, curves of stress versus strain rate, at four different fixed values of the strain, were observed to rise abruptly at a strain rate of about $10^4\,s^{-1}$.  Various investigators have tried to fit these curves by  phenomenological power laws, often ignoring the fact that their predicted stresses would exceed the measured steady-state stresses at higher strain rates.\cite{PTW-03} In his recent review of high-strain-rate deformation, Gray \cite{GRAY-12} asserts that this anomaly and related observations ``support a ... link between substructure evolution and rate-sensitive behavior.''

However, the thermodynamics-based equations of motion [see L15 and Eqs.(\ref{sigmadot}) - (\ref{chidot}) and (\ref{qdef}) below] tell us that this rate-hardening anomaly is a transient phenomenon.  It is controlled by a single rate factor that determines the fraction of the externally applied power that is converted into the energy of new dislocations.  By adding to this conversion factor a single term, linear in the strain rate with a constant coefficient, and possibly identifiable as a grain-size effect, I was able to fit all of the anomalous data and confirm that my results also make sense at higher strain rates.  There is no hint of ``substructure evolution'' in this theory.  But there is testable physical content in it.

Gray's remark about the relevance of substructures is an accurate indication of a general aspect of this situation.  Many different structural changes are observed in connection with deformations of polycrystalline solids.  Examples include twinning, grain-boundary sliding, dynamic recrystallization, formation of cellular dislocation patterns, and the like.  Sometimes it is asserted that these structural changes are the primary mechanisms that cause deformations to take place, sometimes that they are quantitatively relevant but not essential features of complex dynamic processes, and sometimes that they are dynamically irrelevant side effects.  I do not pretend to be able to distinguish between these possibilities in all cases.  But I cannot imagine trying to answer such questions without using effective temperature thermodynamics. 

The formation of dislocation cells is a good example of the relevance of the effective temperature to these theoretical uncertainties.  We know from direct observations that, at least during the early stages of strain hardening, dislocations tend to cluster, forming cellular patterns with low-density interiors and high-density cell walls.  The apparent driving force for this clustering effect is the elastic energy of interaction between dislocations, which adds a term proportional to $-\,\ln\rho$ to the energy $e_D$.  In Eq.~(\ref{Fdef}), this means that both the elastic energy and the entropy decrease logarithmically as the dislocations are brought closer together.  Which of the two effects is dominant depends on the effective temperature $\chi$.  If $\chi$ is small, as it might be for well annealed samples undergoing slow deformation, then the elastic energy is dominant and cells may form.  On the other hand, if $\chi$ becomes large as must happen during later stages of hardening, then the entropy in Eq.(\ref{Fdef}) may be dominant, and cellular structures would no longer be thermodynamically stable.  I cannot claim (yet) to be able to predict the conditions under which cellular dislocation patterns will form, or whether they play any appreciable role in determining the mechanical behaviors of solids.  But the effective temperature surely will have to play a central role in making such predictions.

\section{Equations of Motion}
\label{EOM}

I need now to restate, and comment upon, the thermodynamics-based equations of motion derived in LBL10 and L15.  For the moment, it will be sufficient to consider just a strip of material undergoing uniform simple shear at strain rate $\dot\epsilon$, under a uniform shear stress $\sigma$.  For further simplicity, assume that compressional strains and stresses are negligible.  The dynamical variables are the stress $\sigma$, the dislocation density $\rho$, and the dimensionless effective temperature $\tilde\chi = \chi/e_D$.  

The equation of motion for $\sigma$ is 
\begin{equation}
\label{sigmadot}
\dot\sigma = \mu\,(\dot\epsilon - \dot\epsilon^{pl}),
\end{equation}
where $\mu$ is the shear modulus and $\dot\epsilon$ is the total, elastic plus plastic, strain rate.  I assume that the elastic strains remain small, so that these strain rates are simply additive.

The equation of motion for the dislocation density $\rho$ is 
\begin{equation}
\label{rhodot}
\dot\rho = \kappa_{\rho}\,{\sigma\,\dot\epsilon^{pl}\over \gamma_D}\,\Bigl[1 - {\rho\over \rho_{ss}(\tilde\chi)}\Bigr];~~~~\rho_{ss}(\tilde\chi)= {1\over b^2}\,e^{-1/\tilde\chi};
\end{equation}
where $\gamma_D = e_D/L$ is the dislocation energy per unit length, $L$ is a characteristic dislocation length, and $b$ is an atomic length scale proportional to the length of the Burgers vector. This equation is a statement of energy conservation for the dislocations, consistent with detailed balance.  

The prefactor multiplying the quantity in square brackets in Eq.(\ref{rhodot}), proportional to the input power $\sigma\,\dot\epsilon^{pl}$, is the rate at which new dislocations are formed.  Therefore, the dimensionless coefficient $\kappa_{\rho}$ is the fraction of that power that is converted to the energy of dislocations.  It is the term that becomes linearly strain-rate dependent in order to account for the rate-hardening anomaly mentioned in Sec.~\ref{Teff}. This is one place in the theory where a wide variety of system-specific physical mechanisms may be operative -- e.g. Frank-Read sources, grain-boundary effects, and the like.  We do not need detailed theories of those mechanisms in order to use this equation in predictive calculations.  On the other hand, by measuring $\kappa_{\rho}$ as a function of the strain rate and various material parameters, we should be able to learn a great deal about those mechanisms. 

The second term in the square brackets in Eq.~(\ref{rhodot}) is proportional to the rate at which dislocations are annihilated, in accord with the second law as discussed in L15. Again, we need no specific models of dislocation annihilation in order to write this term; it is determined entirely by the statistical thermodynamic relation between steady-state annihilation and creation rates.  Note that I am using the Boltzmann approximation described in Sec.~\ref{Teff} to write the steady-state dislocation density $\rho_{ss}$ as a function of $\tilde\chi$.  Note also that I am neglecting elastic interactions between dislocations, which would modify the various factors $e_D$ (as well as the equations themselves).   My strategy, as always, is to see how far we can go with the simplest nontrivial assumptions.  

The equation of motion for the effective temperature is
\begin{equation}
\label{chidot}
c_{\it{e\!f\!f}}\,\dot{\tilde \chi} = \sigma\,\dot\epsilon^{pl}\,\Bigl[1 - {\tilde\chi\over \tilde\chi_{ss}}\Bigr].
\end{equation}
This is a statement of the first law of thermodynamics for the configurational subsystem. On the left-hand side of this equation, $c_{\it{e\!f\!f}}= e_D\,\tilde\chi\,\partial S_C/\partial \tilde\chi$ is the effective specific heat.  The second term in the square brackets on the right-hand side is proportional to the rate at which heat flows from the configurational to the thermal (kinetic-vibrational) degrees of freedom.   In writing Eq.~(\ref{chidot}), I have omitted a term proportional to $\dot\rho$ that accounts for energy storage in the form of dislocations, but which often turns out to be negligible.  (See L15.)  Again, for present purposes, I am opting for simplicity.  

So far as I know, Eqs.~(\ref{rhodot}) and (\ref{chidot}) have not appeared in the dislocation theory literature before their publication in LBL10; but they are just statements of the first and second laws of thermodynamics with little specific relevance to dislocations.  Their counterparts in the STZ theory of amorphous plasticity are essentially identical to what is shown here.  Of course, Eq.~(\ref{chidot}) pertains to the effective temperature, which has no counterpart in conventional dislocation theory.  But Eq.~(\ref{rhodot}) replaces the ``storage recovery equation'' of Kocks and Mecking \cite{KOCKS-MECKING-03}, which was argued in LBL10 to be incorrect.  

To close Eqs.~(\ref{sigmadot}) - (\ref{chidot}), we need a dislocation-specific expression for the plastic strain rate $\dot\epsilon^{pl}$. Here is another place where the LBL10 theory differs markedly from the rest of the conventional literature.  But it is here that this theory is most directly a mathematical interpretation of Cottrell's description of the role of dislocations in strain hardening.\cite{COTTRELL-02}  In his words: ``... dislocations are flexible lines, interlinked and entangled, so that the entire system behaves more like a single object of extreme structural complexity and deformability ... a bird's nest.''  He also noted that ``the behavior of the whole system is governed by that of  weakest links.'' I think that he was entirely correct, and that the only negative aspect of his remarks is that he offered them as an explanation of why he thought that strain hardening was an unsolvable physics problem.  In that, I believe he was wrong.  

As argued in LBL10, Cottrell's picture can be made precise by using dimensional arguments that would have been familiar to Orowan and Taylor.  The weak links in Cottrell's bird's nest are the places where dislocations are pinned to each other.  These links are weak in the sense that they can be broken by ordinary thermal fluctuations.  

The LBL10 depinning theory starts with Orowan's dimensional relation between $\dot\epsilon^{pl}$, $\rho$, and the average velocity of the dislocations $v$:
\begin{equation}
\dot\epsilon^{pl}= \rho\,b\,v.
\end{equation}
If a depinned dislocation segment moves a distance of order $\ell \equiv 1/\sqrt{\rho}$ between pinning sites, then $v \sim \ell/\tau_P$, where $1/\tau_P$ is a thermally activated depinning rate given by
\begin{equation}
{1\over \tau_P} = {1\over \tau_0}\,e^{- U_P(\sigma)/k_B T},
\end{equation}  
where $\tau_0$ is a microscopic time, of the order of $10^{-12}$ s.  The activation energy $U_P$ must be a decreasing function of the stress $\sigma$.  Again on dimensional grounds, $\sigma$ must be expressed in units of some physically relevant stress, which seems naturally to be the Taylor stress
\begin{equation}
\label{sigmaT}
\sigma_T(\rho) = \mu\,{b'\over \ell} \equiv \mu_T\,b\,\sqrt{\rho}.
\end{equation}
Here, $b'$ is the dislacement needed for depinning, $b'/\ell$ is the corresponding shear strain, and $\sigma_T$ is the shear stress needed to achieve that strain.  Generally, the reduced modulus $\mu_T$ turns out to be of the order of $0.1~\mu$.  In LBL10, we assumed that 
\begin{equation}
\label{UP}
U_P(\sigma) = k_B\,T_P\,e^{- \sigma/\sigma_T(\rho)},
\end{equation}
where $k_B\,T_P$ is a pinning energy. The exponential function in Eq.~(\ref{UP}) has no special significance; in all applications so far, its argument $\sigma/\sigma_T$ varies by no more than a factor of two or three.  The resulting formula for the dimensionless strain rate $q$ is
\begin{equation}
\label{qdef}
q(\sigma,\rho,\theta) \equiv \tau_0\,\dot\epsilon^{pl} = b\,\sqrt{\rho} \,\exp\,\Bigl[-\,{1\over \theta}\,e^{-\sigma/\sigma_T}\Bigr],
\end{equation}
where $\theta = T/T_P$.  For very small or negative values of the stress, this formula must be antisymmetrized in $\sigma$; but ordinarily that will not be necessary.  

Using the unsymmetrized formula in Eq.(\ref{qdef}), we can solve for $\sigma$ as a function of $q$,  $\rho$, and $\theta$:
\begin{equation}
{\sigma\over\sigma_T(\rho)} = \ln\Bigl({1\over \theta}\Bigr) - \ln\Bigl[\ln\Bigl({b^2\,\rho\over q}\Bigr)\Bigr]\equiv \nu(\rho,q,\theta).
\end{equation}
The quantity $\nu(\rho,q,\theta)$ is a very slowly varying function of its arguments, consistent with the observation that the Taylor stress is a good approximation to the true stress in most circumstances, and also with the observation that steady-state rate hardening is generally quite slow.  It is this formula, with $\rho = \rho_{ss}$, that produces agreement with experiment for steady-state stresses over fifteen decades of strain rate as shown in LBL10 and mentioned here in Sec.~\ref{Teff}.  Conversely, this formula tells us that the strain rate $q(\sigma,\rho,\theta)$ is a highly sensitive function of its arguments, especially -- for present purposes -- the dimensionless temperature $\theta$.

\section{Thermal Softening}
\label{Thermal}

A key issue that cannot be resolved without a first-principles theory of dislocation dynamics is the role of ordinary thermal fluctuations in enabling failure modes such as shear localization and fracture.  Clearly, the formation of dislocations is a hardening mechanism.  The occurrence of shear localization implies that there must be a competing softening mechanism, which generally has been assumed to be a thermal instability.  A local increase in strain rate produces a local increase in temperature that, in turn, locally softens the material and thus further increases the strain rate.  

Many shear-banding analyses have appeared in the literature in recent decades.  (For example, see \cite{WRIGHT-02, ASL-12}.)  All of these use phenomenological models of strain hardening and thermal softening that I find unrealistic for reasons stated in Sec.~\ref{Teff}.  Moreover, the assumption that thermal softening is the primary driver of localization instabilities has been  challenged by Rittel and coworkers \cite{RITTELetal-10,RITTELetal-12}, who show credible experimental evidence that dynamic recrystallization (DRX) occurs in the neighborhood of adiabatic shear bands {\it before} the main, heat generating shear band appears.  He argues, therefore, that the primary softening mechanism may be DRX rather than a conventional heating process.

As a first step toward a study of these issues, it would be straightforward to use the equations of motion shown in Sec.~\ref{EOM}, supplemented by an equation of motion for the dimensionless temperature $\theta$, in a shear-banding theory like that described by Manning et al. \cite{MANNINGetal-SHEARBANDS-08} for amorphous materials.  That is, we could consider an infinitely long strip of width $W$ in the $x$-$y$ plane, driven in simple shear by constraining it to be fixed along the $x$ axis, $y = 0$, and to be moving at constant speed $v_W$ along its upper edge at $y = W$.  As in  \cite{MANNINGetal-SHEARBANDS-08}, we could neglect dilational deformations, and assume that the shear stress $\sigma$ is a spatially uniform function of only the time $t$. Then we could solve for variations of $\dot\epsilon^{pl}$, $\rho$, $\tilde\chi$ and $\theta$ as functions of $y$ and $t$, and look for instabilities in which the plastic strain rate becomes localized in a band near some position along the $y$ axis.  This model still does not include the shape changes and stress concentrations that were centrally important in the study of fracture toughness in metallic glasses by Rycroft and Bouchbinder \cite{RYCROFT-EB-12}. However, even this simplified model seems to me to be more complicated than is needed here in order to gain a first understanding of what is happening. 

Suppose that our strip consists of just two parallel zones, one occupying the region $0 < y < \alpha\, W$ and the other occupying $\alpha\,W < y < W$.  Denote these zones by the index $ n = 0,\,1$.  The total strain rate is $\dot\epsilon = v_W/W$. With the time $t$ replaced by total strain $\epsilon$  as the independent variable, and $Q \equiv\tau_0\,\dot\epsilon$, the condition of continuity at $y = \alpha \,W$ becomes an equation of motion for the spatially uniform stress:
\begin{equation}
{d\,\sigma\over d\epsilon} = \mu\,\Bigl[ 1 - \alpha\,{q_0\over Q} - (1-\alpha)\,{q_1\over Q}\Bigr],
\end{equation}
where
\begin{equation}
\label{qdef2}
q_n = \sqrt{\tilde\rho_n} \,\exp\,\Bigl[-\,{1\over \theta_n}\,e^{-\sigma/\sigma_T(\tilde\rho_n)}\Bigr];
\end{equation}
and
\begin{equation}
\label{sigmaT2}
\sigma_T(\tilde\rho_n)= \mu_T\,\sqrt{\tilde\rho_n};~~~~\tilde\rho_n \equiv b^2\,\rho_n.
\end{equation}

\begin{figure}[here]
\centering \epsfig{width=.45\textwidth,file=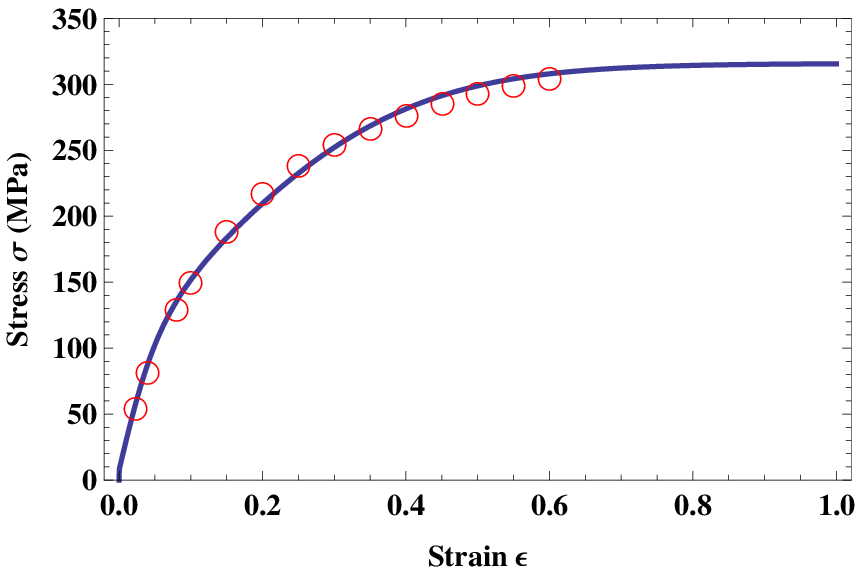} \caption{Stress versus strain for the small strain rate, $\dot\epsilon = 0.002~s^{-1}$.  The open circles are the experimental points shown in LBL10, Fig.~2. }\label{Fig1}
\end{figure}
\begin{figure}[here]
\centering \epsfig{width=.45\textwidth,file=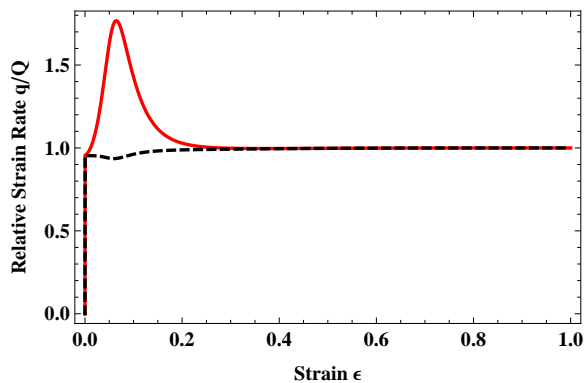} \caption{Relative strain rates $q_n/Q$ for $\dot\epsilon = 0.002 ~s^{-1}$. The upper solid red curve is for the narrow zone 0; the lower dashed black curve is for zone 1. The fractional width of the narrow zone is $\alpha = 0.05$. }\label{Fig2}
\end{figure}

Eq.~(\ref{rhodot}) becomes
\begin{equation}
\label{rhodot2}
{d\,\tilde\rho_n\over d\epsilon} = \kappa_1\,{\sigma\,q_n\over \nu_n^2\,\mu_T\,Q}\, \Bigl[1 - {\tilde\rho_n\over e^{-1/\tilde\chi_n}}\Bigr] .
\end{equation}
Here, 
\begin{equation} 
\nu_n = \sigma/\sigma_T(\tilde\rho_n)= \ln\Bigl({1\over \theta_n}\Bigr) - \ln\Bigl[\ln\Bigl({\tilde\rho_n\over q_n}\Bigr)\Bigr].
\end{equation}
The prefactor in Eq.(\ref{rhodot2}), proportional to $\kappa_1$, was shown in LBL10 and L15 to be the value of $\kappa_{\rho}$ in Eq.(\ref{rhodot}) needed to describe the onset of hardening.  $\kappa_1$ is a dimensionless constant of the order of unity.  Equation~(\ref{chidot}), the equation of motion for $\tilde\chi$, becomes
\begin{equation}
\label{chidot2}
{d\,\tilde\chi_n\over d\epsilon} = \kappa\,{\sigma\,q_n\over \mu_T\,Q}\,\Bigl[ 1 - {\tilde\chi_n\over \tilde\chi_{ss}}\Bigr],
\end{equation}
where $\kappa$ is a dimensionless constant inversely proportional to $c_{\it{e\!f\!f}}$, and also of the order of unity. 

It remains to write equations of motion for the temperatures $\theta_n$.  For the two-zone model, these have the form:
\begin{equation}
\label{dottheta0}
{d\theta_0\over d\epsilon} = K\,{\sigma\,q_0\over \mu_T\,Q} + {K_1\over Q}\,(\theta_1 - \theta_0) - {K_2\over Q}\,(\theta_0 - \theta_a);
\end{equation} 
and
\begin{equation}
\label{dottheta1}
{d\theta_1\over d\epsilon} = K\,{\sigma\,q_1\over \mu_T\,Q} - {K_1\over Q}\,(\theta_1 - \theta_0)- {K_2\over Q}\,(\theta_1 - \theta_a).
\end{equation}
The first terms on the right-hand sides of these equations are the rates at which heat is generated in the two zones.  The factors $\mu_T^{-1}$ have been inserted for dimensional consistency.  The second terms on the right-hand sides are the rates at which heat flows between the zones. The last terms are the rates at which the zone temperatures relax toward the ambient temperature $T_a$, where $\theta_a = T_a/T_P$. 

\begin{figure}[here]
\centering \epsfig{width=.45\textwidth,file=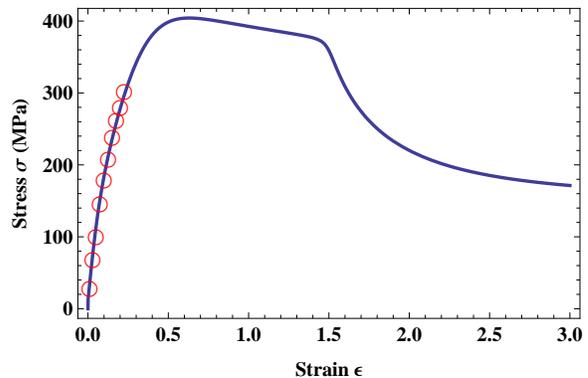} \caption{Stress versus strain for the large strain rate $\dot\epsilon = 2000~s^{-1}$ .  The open circles are the experimental points shown in LBL10, Fig.~2.}\label{Fig3}
\end{figure}

\begin{figure}[here]
\centering \epsfig{width=.45\textwidth,file=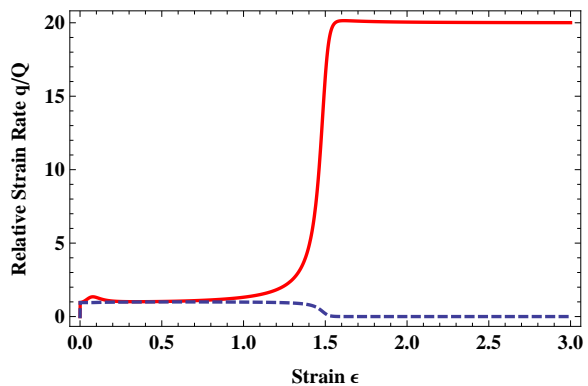} \caption{Relative strain rates $q_n/Q$ for $\dot\epsilon = 2000~s^{-1}$. The upper solid red curve is for the narrow zone 0; the lower dashed black curve is for zone 1.  The fractional width of the narrow zone is $\alpha = 0.05$; thus the fact that $q/Q \to 20$ for this zone means that it carries essentially all of the strain rate after the band forms. }\label{Fig4}
\end{figure}

\begin{figure}[here]
\centering \epsfig{width=.45\textwidth,file=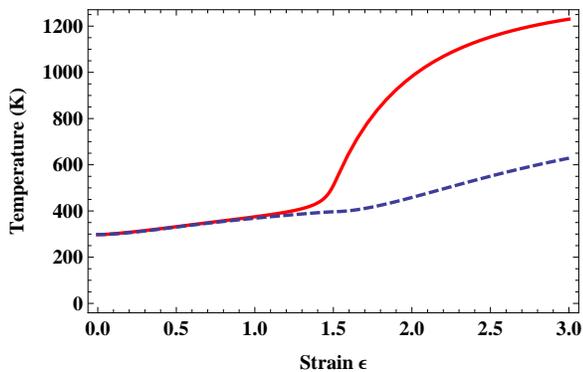} \caption{Temperature $T$ versus strain for $\dot\epsilon = 2000~s^{-1}$. The upper solid red curve is for the narrow zone 0; the lower dashed black curve is for zone 1. }\label{Fig5}
\end{figure}

Solutions of these equations are shown here in Figs. \ref{Fig1} - \ref{Fig5} for the same two cases of room-temperature Cu whose stress-strain curves are shown in LBL10  Fig.~2.  The only difference between these two cases is that they are measurements at very different strain rates, $\dot\epsilon = 0.002~s^{-1}~~{\rm and}~~ 2000~s^{-1}$; I use them to explore the strain-rate dependence of the banding instability.  Apart from the thermal coefficients $K$, $K_1$ and $K_2$ in Eqs.~(\ref{dottheta0}) and (\ref{dottheta1}), all material parameters are the same as in LBL10: $T_P = 40800\,K$; $T_a= 298\,K$, which is also the initial temperature for both zones; $\mu_T = 1600\,\,{\rm GPa}$; $\mu = 31\,\mu_T$; $\tilde\chi_{ss} = 0.25$; $\tau_0 = 10^{-12}\,s$; $\kappa = 11.2$; and $\kappa_1 = 3.1$.  I have arbitrarily chosen $K = 0.01$ and $K_1 = K_2 = 10^{-9}$ (so as to be roughly comparable in magnitude to the larger values of $Q$). The initial value of $\tilde\rho_n$ is $10^{-5}$ in all cases.  In order to trigger an instability, I have used slightly different initial values of $\tilde\chi$ for the two zones, specifically $0.16$ for zone 0 and $0.18$ for zone 1.  Finally, I have chosen $\alpha = 0.05$, so that zone 0 is very narrow.  It is the zone that becomes a shear band at the higher strain rate.

At the smaller strain rate for which results are shown in Figs. \ref{Fig1} - \ref{Fig2}, there is no indication of any shear-banding instability.  The stress-strain curve in Fig.\ref{Fig1} is indistinguishable from the experimental data in the initial hardening regime.  Figure \ref{Fig2} shows that, after an initial transient in the relative strain rate $q_0(\epsilon)/Q$ for the narrow zone, the strain rate becomes uniform across the system.  The heat flow described by the second and third terms on the right-hand sides of Eqs.~(\ref{dottheta0}) and (\ref{dottheta1}) is fast enough to quench the instability.  The temperatures of both zones remain at their initial values throughout the process. 

The fast system described in Figs. \ref{Fig3} - \ref{Fig5} is more interesting.  The stress-strain curve in Fig.\ref{Fig3} looks similar to curves of this kind shown by Rittel \cite{RITTELetal-10,RITTELetal-12}.  It has a smooth peak at $\epsilon \cong 0.5$ where an instability sets in, and a sharper drop at $\epsilon \cong 1.3$ where the system ``breaks'' in the sense that, as seen in Fig.\ref{Fig4}, almost all of the strain rate is suddenly shifted to the narrow zone 0. The zone temperatures shown in Fig.\ref{Fig5} become distinctly different from each other as this happens. Note, however, that this thermal effect sets in at $\epsilon \cong 0.7$, well before the band forms.  Whether or not this thermal precursor is related to the early occurrence of DRX remains to be seen. 

The results of these numerical experiments  reflect the nonlinear nature of the thermal softening instability governed by the preceding equations, especially by the strong $\theta$ dependence of the strain rate shown in Eqs.(\ref{qdef}) and (\ref{qdef2}).  For example, using the same material parameters and initial conditions, and varying the imposed strain rate, I numerically find the first signs of the instability appearing at $\dot\epsilon \cong 560~s^{-1}$.  Here, however, shear localization sets in very slowly and does not occur visibly until the system has reached a strain of about $\epsilon \cong 26$, far larger than would be observed in realistic experiments.  The onset of instability is also quite sensitive to small changes in other parameters, especially the thermal transport coefficients $K$, $K_1$, and $K_2$.  However, I see no reason to think that a mathematically sharp instability occurs at any nonzero strain rate.  

The most serious weakness of this toy model is that it does not tell us what physical mechanisms determine the width or structure of shear bands.  The band width $\alpha$ must be chosen {\it a priori} and, as used here, plays no role in determining the heat flow between the band and the neighboring material.  The latter problem might be remedied by going to a continuum description of the spatial dependence of the internal variables, in analogy to the calculations of Manning et al. \cite{MANNINGetal-SHEARBANDS-08} for amorphous systems.  Such an analysis may be a next step in this project.

\section{Concluding Remarks}

The dislocation theory presented here, unlike much of the conventional literature in this field, is fully consistent with the laws of thermodynamics and basic physical principles of symmetry, energy conservation, and the like. It is couched in terms of properly defined state variables as discussed in \cite{BL-I-II-III-09}.  For example, it does not artificially distinguish between ``mobile'' and ``immobile'' dislocations, nor does it use a plastic strain field and its associated reference state as if these were physically meaningful concepts in irreversible systems.  It is much simpler than conventional theories, consisting of just a few equations of motion and a comparably small number of physically meaningful, material-specific parameters.  There are no phenomenological power-law fits to experimental data.  And yet, this theory has proven capable of explaining quantitatively a broad range of previously unexplained behaviors, as described in LBL10 and L15.  

In my opinion, the main weakness of this simple theory, so far, is that it is {\it too} simple.  It seems essential to find its limits of validity.  One way to do that will be to repeat the analyses of LBL10 and L15 using other sets of experimental data, measured for different kinds of materials under different driving conditions.  Clear discrepancies will indicate missing or incorrect theoretical ingredients. A strategy especially relevant to the present paper would be to use the Rycroft-Bouchbinder techniques \cite{RYCROFT-EB-12} with this dislocation theory to compute fracture toughness, or to look in detail at Rittel's observations. \cite{RITTELetal-10,RITTELetal-12}; but that strategy would have to be coordinated with new experimental obsrvations.  

More generally, we need to explore the ways in which this theory is, or is not, able to interpret collective dynamic behaviors of dislocations such as those described in the monumental review by Ananthakrishna \cite{ANANTHAKRISHNA-07}.  In his presentation, Ananthakrishna is always careful to couch his mathematics in terms of physically well defined state variables (e.g. dislocation densities), and he uses those variables in statistically sensible Langevin or Fokker-Planck equations.  In my opinion, he does not take the concept of entropy seriously enough, and therefore runs into trouble evaluating the noise strengths that necessarily enter those formulations.  Nevertheless, his discussions of various kinds of slip bands and propagating modes such as the Portevin-Le Chatelier effect seem to me to be very interesting.  They may lie outside the range of the present theory.  If so, I would much like to understand why. 

\begin{acknowledgments}

This research  was supported in part by the U.S. Department of Energy, Office of Basic Energy Sciences, Materials Science and Engineering Division, DE-AC05-00OR-22725, through a subcontract from Oak Ridge National Laboratory.

\end{acknowledgments}

\end{document}